\documentclass[aip,graphicx]{revtex4-1}

\draft % marks overfull lines with a black rule on the right

\begin{document}

%\preprint{}

\title{Classical novae and type I X-ray bursts: challenges for the 21st century} %Title of paper


\author{A. Parikh}
\email[]{anuj.r.parikh@upc.edu}
%\thanks{}

\author{J. Jos\'e}
\email[]{jordi.jose@upc.edu}

\author{G. Sala}
\email[]{gloria.sala@upc.edu}

\affiliation{Departament de F\'isica i Enginyeria Nuclear, Universitat Polit\`ecnica de Catalunya (EUETIB), E-08036 Barcelona, Spain}
\affiliation{Institut d'Estudis Espacials de Catalunya (IEEC), E-08034 Barcelona, Spain}
%\author{}
%\thanks{}
%\affiliation{}


\date{\today}

\begin{abstract}
Classical nova explosions and type I X-ray bursts are the most frequent types of thermonuclear
stellar explosions in the Galaxy. Both phenomena arise from thermonuclear ignition in the envelopes
of accreting compact objects in close binary star systems. Detailed observations of these
events have stimulated numerous studies in theoretical astrophysics and experimental nuclear
physics. We discuss observational features of these phenomena and theoretical efforts to better understand the energy production and nucleosynthesis in these
explosions.   We also examine and summarize studies directed at identifying nuclear physics quantities with uncertainties that significantly affect model predictions.
\end{abstract}

\pacs{26.30.Ca, 26.50.+x, 97.80.Gm, 97.80.Jp}% insert suggested PACS numbers in braces on next line

\maketitle %\maketitle must follow title, authors, abstract and \pacs

%\section{}
%\label{}
%\subsection{}
%\subsubsection{}

\section{Introduction}
\label{intro}

 {\it``The only true wisdom is in knowing you know nothing."} - Socrates\\

The first systematic registry of novae was initiated around 200 BCE by officials of the Chinese imperial court (see Duerbeck\cite{gsdue08} for a list of observed novae up to 1604). From the nineteenth century until the 1950s, careful monitoring of the sky with photographic plates led to the discovery of many events.  Afterwards and until the 1970s, novae were mainly found through surveys using objective prisms. Since the 1980s, almost all novae have been discovered by amateurs, first using binoculars and later through the analysis of CCD images.  Note that the term \emph{nova} was originally used to denote all bright, star-like phenomena that suddenly appeared in the night sky and gradually faded with time.  It was only in the 1920s and
1930s, through dedicated studies examining the distances to different nova events (distinguishing
those of Galactic origin from those at much greater distances) that the distinct terms \emph{supernova} and
\emph{nova} were introduced to differentiate between the intrinsically brighter supernova events and the
dimmer nova events.  Several hundred Galactic novae have been discovered
to date, with $\approx$5 events discovered per year. They are characterized by peak luminosities of
$\approx$10$^{4}$ --10$^{5}$ L$_{\odot}$, light curves of $\approx$days to months in duration, and mass ejection into the interstellar
medium of $\approx$10$^{-4}$ -- 10$^{-5}$ M$_{\odot}$ per event.  The recurrence time is expected to be $\approx$10$^{4}$ -- 10$^{5}$ years, although the subclass of recurrent novae have recurrence times of only years or decades.

The first X-ray burst was identified in 1975\cite{gsgri76} from a previously-known X-ray source, 4U 1820-30.  [Note that most X-ray sources are named using letters 
from the satellites that discovered them (e.g., 
4U stands for the 4th catalogue of the satellite Uhuru, the first satellite dedicated to X-ray astronomy), and numbers corresponding to their coordinates 
in Right Ascension (e.g., 1820 stands for 18h 20min) and Declination 
(--30 deg) in the sky.  They may also be named after the 
constellation where they are located and the order of discovery.  As a result, a source may have more than one name; e.g., 4U 1820-30 is also known as Sgr X-4.]
A similar episode had been observed
in 1969 from the source Cen X-4\cite{gsbel72}, 
but the authors related the observed feature to an accretion event and 
it was not recognized as a new type of source until 1976 \cite{gsbel76, gskul09}. Cen X-4 is still the nearest-known 
XRB source (at $\approx$1 kpc) and has yielded the brightest burst ever recorded 
($\approx$50~Crab = 50 $\times$ 2.4$\times10^{-8}$ erg s$^{-1}$ cm$^{-2}$ in the 2--10 keV band).  Roughly 100 bursting systems have been identified in
the Galaxy, with light curves of $\approx$10 -- 100 s in duration, recurrence periods of $\approx$hours to days and
similar peak luminosities to those of classical novae. Calculations indicate that radiative winds
generated during some bursts may eject material; studies are ongoing to examine the viability of
detecting absorption features arising from this material.  

Both classical nova explosions (CN) and type I X-ray bursts (XRBs) arise from thermonuclear runaways
within the accreted envelopes of compact objects in close binary systems, with orbital periods
often less than 15 hours. Generally in a classical nova, H-rich material is transferred from a
low mass main sequence or red giant star onto the surface of a white dwarf star. In a type I X-ray
burst, a neutron star interacts with a similar low mass companion star; observations are
consistent with the accretion of material enriched in H, He or both. Typical accretion rates in
these events range from 10$^{-10}$ -- 10$^{-9}$ M$_{\odot}$/yr, although the range and implications of different
accretion rates (resulting in e.g., stable or marginally stable burning for an accreting neutron star) are still under investigation. As accretion proceeds, the envelope is gradually compressed and
becomes degenerate. The temperature of the envelope increases, creating conditions favorable to
the ignition of the accreted fuel through nuclear reactions. These reactions, once initiated, drive
further reactions, leading to the thermonuclear runaway and the corresponding explosion. Note that
the degeneracy of the envelope is lifted as the temperature increases at early times. For example,
for the accretion of material with $Z/A\approx0.5$ , degeneracy is lifted at T$\approx$30 MK or $\approx$300 MK for
a white dwarf or neutron star envelope, respectively. The difference between CN and XRBs from
the viewpoint of mass ejection arises from how in the latter case the necessary escape velocities are
never achieved. 

In this article, we will focus on standard models of CN and models of XRBs that occur in envelopes containing substantial H and He.  We briefly survey the evolution and state-of-the-art of our knowledge of these phenomena.   We refer the reader to, e.g., Bode and Evans\cite{Bod08} and Jos\'e and Hernanz\cite{Jos07rev} (for CN) and, e.g., Lewin et al.\cite{Lew93}, Strohmayer and Bildsten\cite{Str06} and Parikh et al.\cite{Par13rev} (for XRBs) for more extensive reviews.  The framework of this article reflects the approach requested by the editors, namely, to address less what is known, and more what is as yet unknown by discussing questions that still need answering and which methods are the most powerful for doing so.  In this vein, we present below four headings regarding needs to better constrain predictions from models of classical novae and type I X-ray bursts.  For each heading we provide support and background information as appropriate.  We certainly do not claim to have summarized all the varied outstanding problems and challenges that remain for these phenomena.  (For example, we do not discuss the unexplained oscillations observed in the light curves of XRBs\cite{Str96,Str06,Wat12} or in the soft X-ray light curves from novae\cite{gsdra03,gsdob10,gsness11}, both of which may be indicative of a confined radiating region.)  Nonetheless, the issues discussed below represent major tasks or obstacles that need to be addressed to improve our understanding of these thermonuclear explosions.

\section{Are spherically symmetric models still needed?}

Different methods and numerical approaches have been used to determine the nucleosynthesis accompanying 
novae and X-ray bursts.  One category of models relies on parameterized one-zone (or multi-zone) calculations 
(e.g., Refs. \cite{HT82, Wan99, WT90, Wie86} for models of CN, and Refs. \cite{FHM81, Sch99, Wal81, Han83, Has83, Koi99, Koi04, Sch01} for XRB models).
These prescriptions (coined by some authors as 0-D!) relate the thermodynamic history of the compact object's accreted envelope with the time evolution of the 
temperature and density in a single layer (often the innermost) or multiple envelope layers.  These temperature-density-time profiles are frequently 
determined by means of semianalytical models or extracted from 1-D hydrodynamic 
calculations. While representing an extreme oversimplification of the thermodynamic conditions characterizing the envelope,
the approach has been extensively used as a way to overcome the otherwise computationally prohibitive calculations that a purely hydrodynamic
approach would require. Recently, these techniques have been adopted in sensitivity studies of nucleosynthesis in CN and XRBs to variations in rates of nuclear interactions\cite{Ili02, Amt06, Rob06, Par08, Par09} (see Section \ref{sens}).

To date, state-of-the-art nucleosynthesis calculations rely exclusively on 1-D hydrodynamic models (e.g., Refs. \cite{JH98, PK95, Sha10, Sta09, Sta98, Yar05} 
for nucleosynthesis in CN, and Refs. \cite{Woo04, Fis08, Jos10, WW84, 
Wal82, Taa93, Taa96}, for nucleosynthesis in XRBs). The underlying assumption in all of these models is, obviously, spherical symmetry. The implication is that the explosion is modeled as occurring uniformly (and simultaneously) over a spherical shell. In sharp contrast, 
these thermonuclear runaways are expected to originate from point-like ignitions. 
Clearly, detailed nucleosynthesis models require multidimensional hydrodynamic simulations, but this will only be feasible when sufficient computational power is available to model all relevant details of these explosions (see Section \ref{multiD}). 

\subsection{Classical novae}

Perhaps the first discussion of the physical
mechanism powering nova explosions appears in Newton's {\it Principia Mathematica}:
{\it "So fixed stars, that have been gradually wasted by the light and vapors emitted from them
for a long time, may be recruited by comets that fall upon them; and from this fresh supply of
new fuel those old stars, acquiring new splendor, may pass for new stars".} 
It is worth noting that the revitalization of an old star through the fresh supply of new fuel (although not
by comets!) is at the core of the {\it thermonuclear runaway model}, in which mass transfer in a 
binary system plays the required role.

The underlying physics of the nova phenomenon was exposed, in part, through a number
of observational breakthroughs, including:
\begin{itemize}
\item the first optical spectroscopic analysis of nova T CrB 1866 \cite{HM1866}
\item discovery in 1901 of spectroscopic features at 3869 and 3968 \AA (later identified as Ne III lines) in the spectra of GK Per
 \cite{Sid01a,Sid01b}, suggesting the existence of different nova types (although the first
  simulations of novae hosting oxygen-neon (ONe) white dwarfs were not performed until the mid-1980s \cite{Sta86})
\item the explanation of the observed spectral features as due to ejection of a shell from a star \cite{Pic1895}
\item the link between the minimum in the DQ Her light curve and episodic dust formation \cite{SM39}
\item discovery of the binary nature of DQ Her \cite{Wal54}
\item systematic studies of novae revealing that binarity is a common property of most
cataclysmic variables (novae, in particular \cite{Kra64a,Kra64b,San49,Joy54})
\end{itemize}
While the observational picture was firmly established on the basis of ejection from the
surface of a star, explanations of the physics behind the burst were not offered until the middle of the twentieth century. Indeed, its thermonuclear
origin was first theorized by Schatzmann \cite{Sch49,Sch51}, although incorrectly interpreted as due to nuclear fusion reactions 
involving $^3$He. Other notable theoretical contributions were 
made in the late 1950s \cite{GL57,Cam59}; attempts to compute the explosion in a hydrodynamic frame
were published a decade later \cite{GW67,Ros68,Spa69}. The idea that CNO enhancement is critical for the energetics of the explosion was first 
proposed by Starrfield et al. \cite{Sta71a,Sta71b}.  

Several groups have analyzed in detail the nucleosynthesis accompanying nova explosions \cite{Sta98,JH98,KP97,Den13}
by coupling nuclear reaction networks, containing about a hundred species and a few hundred nuclear interactions, directly
to 1-D hydrodynamic models. While these models have been successful in reproducing the gross observational features of nova
outbursts (e.g., nucleosynthesis, peak luminosities L$_{peak}$, light curves), the exact amount of material ejected by the explosion is
still a matter of debate. A next generation of models with state-of-the-art input physics, methods to treat rotation, and better techniques to tackle convective transport (i.e., based on results from 3-D models), will be needed to shed light into these matters.

\subsection{Type I X-ray bursts}

The first estimates of the amount of nuclear energy that may be released from the fusion of H-rich material in envelopes
accreted by neutron stars were made by Rosenbluth et al. \cite{Ros73}. The scenario was revisited shortly afterwards 
in studies\cite{HvH75,vHH74} that revealed that nuclear fusion may trigger thermonuclear runaways. The thermonuclear origin
of type I X-ray bursts, resulting from unstable nuclear fusion on the surfaces of neutron stars, was first suggested
by Woosley and Taam \cite{WT76}(for bursts driven through He-burning) and Maraschi and Cavaliere \cite{MC77} (for bursts driven through H-burning).  The mechanism was further defined through a number of increasingly detailed simulations \cite{FHM81,Pac83,Jos78,TP79,CJ80,JL80,ET80,Taa80,Taa81,Taa82,AJ82}.
The gross observational features of type I XRBs were succesfully reproduced by early studies: using dimensional analysis, Joss \cite{Jos77} and Lamb and Lamb \cite{LL78} inferred recurrence periods of about 10 000 s,
accreted envelope masses of $\approx 10^{21}$ g, and an overall energy release of $10^{39}$ erg per burst (assumed to be driven through He burning).
Other estimates of burst properties included peak luminosities L$_{peak} \approx 10^{38}$ erg s$^{-1}$, light curve 
rise times of $\approx 0.1$ s, and decay times of $\approx 10$ s, on the basis of nuclear energy transport from the deepest envelope 
layers to the outermost region. Another important observational constraint matched by thermonuclear models of type I
XRBs was the so-called {\it $\alpha$} parameter, or ratio of persistent over burst luminosities.

As shown in the pioneering work of Joss \cite{Jos78}, the key parameters in the modeling of type I X-ray bursts are the 
mass accretion rate, the neutron star mass and central temperature (which in turn determines the initial surface luminosity of the neutron star) as well as the metallicity of the accreted material. This work confirmed the estimates obtained from previous dimensional analysis studies. 
Furthermore, it showed that all the accreted fuel is essentially consumed during the burst (processing H-rich material into mostly
Fe-peak elements), due to an efficient CNO-breakout, in sharp contrast with the much more limited nuclear activity exhibited by
classical novae. As well, most of the energy released during an X-ray burst is emitted as X-rays from the star's photosphere. 

Detailed nucleosynthesis studies under the characteristic temperatures and densities reached in neutron star envelopes
(with peak values around $10^9$K and $10^6$ g cm$^{-3}$) require huge nuclear reaction networks, with hundreds of isotopes and 
thousands of nuclear interactions. Initially, this was only feasible in the framework of one-zone models \cite{Koi99, Sch99, Sch01, Koi04}.
These pioneering studies revealed that the main nuclear reaction flow is driven by the rp-process (rapid proton-captures and $\beta^+$-decays), the
3$\alpha$-reaction, and the $\alpha$p-process (a sequence of ($\alpha$, p) and (p, $\gamma$) reactions). 
Only recently has it been possible to use detailed nuclear reaction networks in a purely hydrodynamic framework (1-D -- see Refs. \cite{Woo04, Fis08, Jos10}
and references therein). The extension of the nuclear activity in XRBs is still a matter of debate, since recent experimental studies \cite{Elo09} 
have shown difficulties in reaching the SnSbTe-mass region, previously identified as the likely nucleosynthesis endpoint\cite{Sch01}.
Additional difficulties in the modeling arose from the discovery of highly magnetized neutron stars (with B $\geq 10^{12}$ G),
in which mass accretion from the stellar companion is expected to be funnelled onto the neutron star magnetic poles, enhancing the local 
accretion rates in those spots by $\approx$ three orders of magnitude \cite{JL80}. A number of different ignition regimes for specific ranges of
mass accretion rates have been inferred for accretion of solar-like material \cite{FHM81,FL87}.

General relativistic corrections to calculations performed using a Newtonian framework were first incorporated into models in the 1980s \cite{TP79,CJ80,Taa80,AJ82}.
In short, their effect is a net reduction of the expected peak luminosities and an enhancement of the recurrence times by a factor of $1+z$, with
$z$ being the gravitational redshift of the neutron star.

While the modeling of type I XRBs resulting from the combined burning of H and He has been emphasized recently (in part due to the interest of experimentalists in constraining the associated nucleosynthesis), more work needs to be done to explore the nature of superbursts (roughly 1000 times more energetic than type I XRBs, with recurrence times on the order of a year)\cite{Cum01,Kee11,Kee12} and bursts intermediate in both energy and duration to typical type I XRBs and superbursts\cite{Zan02,Zan05}.  Moreover, 1-D models are still needed
to resolve possible discrepancies in the extent of the nuclear activity for low-metallicity environments (e.g., Refs. \cite{Jos10, Fis08}), the possible nuclear origin of double- or multiple-peaked bursts\cite{Fis04,gsbs06a}, and the possible contribution of type I X-ray bursts to Galactic abundances through radiation-driven winds\cite{Kat83,Pac83,Nob94,Wei06a, Zha11}.

\section{Toward multidimensional models}
\label{multiD}

The assumption of spherical symmetry has been adopted in the vast majority of models of CN and XRBs, with which gross observable features of these phenomena have been reproduced.  It is clear, however, that this assumption excludes details associated with the manner in which a thermonuclear runaway initiates (presumably as a point-like ignition) and propagates. Flames may propagate supersonically (detonations) or subsonically (deflagrations). In both CN 
and XRBs, burning fronts are expected to propagate subsonically. Such deflagration models are 
more difficult to compute than detonation models.  Indeed, standard 
compressible hydrodynamics codes usually fail when applied to deflagrations because of the long 
integration times required.  This is a major reason why more multidimensional models of, e.g., 
Type Ia supernovae\cite{Hil13} (in which at least part of the explosion proceeds as a detonation, according to current models) have been published than of classical novae or type I X-ray bursts.

The first study of localized runaways in degenerate envelopes, involving
white dwarfs or neutron stars, was carried out by Shara \cite{Sha82} on the basis of 
semianalytical models. He suggested that heat transport was too inefficient to spread a localized flame
(i.e., the diffusively propagated burning wave may require
tens of years to extend along the entire white dwarf surface), and concluded that localized, {\it volcanic-like}
eruptions were likely to occur. Unfortunately, while this analysis did consider
radiative and conductive energy transport, it ignored the major role 
played by convection on the propagation of the burning front.
Indeed, as soon as superadiabatic gradients are established, 
macroscopic mass elements are exchanged between hotter and cooler regions of 
the envelope through convective transport. These mass elements ultimately dissolve in the environment, releasing their excess heat.
Because of its complexity, the treatment of heat transfer in convective zones is often tackled by means 
of phenomenological approaches (i.e., mixing-length theory). Unfortunately, 
convection is a truly multidimensional process that cannot be reliably modeled under the assumption of spherical symmetry. 

The importance of multidimensional effects in explosions occurring in 
thin stellar shells was revisited by Fryxell and Woosley \cite{FW82}. The study
concluded that the most likely scenario 
involves runaways propagated by small-scale turbulence in a deflagrative
(subsonic) regime, leading to the horizontal spread of the front at typical velocities of 
$v_{\rm  def} \sim  10^4$ cm  s$^{-1}$ (for CN) and 
$v_{\rm  def} \sim  5 \times 10^6$ cm  s$^{-1}$ (for XRBs).

The first attempts to address the importance of multidimensional
effects on nova explosions in a truly hydrodynamic framework were performed 
by Shankar et al. \cite{SAF92, SA94}. To this end, a 1.25 $M_\odot$ accreting white dwarf 
was evolved in spherical symmetry, and subsequently mapped into a 2-D 
domain.  Because of computational limitations, only a small section of the star, 
a spherical-polar shell of 25 km$\times$60 km, was considered. 
The explosive stage was then followed in
2-D with the explicit, Eulerian code {\it PROMETHEUS}.  Unfortunately, 
the subsonic nature of the problem, coupled with timestep limitations
because of the explicit nature of the code, posed severe
constraints on the simulation. Indeed, the authors were forced to adopt very extreme
conditions that resulted in unrealistic detonation waves propagating throughout
the accreted envelope. Similar problems were encountered in the first multidimensional
simulations of thermonuclear explosions in neutron star envelopes \cite{Zin01} (note
that although this study is often referred to as the first multidimensional simulation of an XRB, it solely addressed helium detonations on neutron stars). 
The work of Shankar et al. revealed  that instantaneous, local
temperature fluctuations can induce Rayleigh-Taylor instabilities, whose
rapid rise and subsequent expansion can cool
the hot material and halt the lateral spread of the burning
front.  This would favor local, volcanic-like eruptions.
Nonetheless, a full hydrodynamic simulation
performed under realistic conditions was still needed to shed further light on this issue.

Improved simulations were published shortly
after by Glasner et al.  \cite{GL95, GLT97}.  These were performed in two dimensions with the code  {\it VULCAN}, an arbitrarily Lagrangian Eulerian (ALE) hydrodynamic code capable of handling both explicit and implicit steps.  A $1 M_\odot$ accreting carbon-oxygen (CO) white dwarf was first evolved using a 1-D hydrodynamic code, and then 
mapped into a 2-D domain (again, because of computational limitations, only a thin slice of the star, $0.1\pi$ rad, was considered).
These new simulations showed that the thermonuclear runaway initiates
as  a myriad of irregular, localized eruptions that appear close to the
envelope base, each surviving for
only a few seconds.  Turbulent diffusion efficiently dissipates any local burning around the core, 
and hence the flame must eventually spread along the entire envelope.
Large convective eddies, extending up to $2/3$ of the envelope height (with typical velocities  
$v_{\rm conv} \sim 10^7$  cm s$^{-1}$) were found.  
The core-envelope interface appears to be 
convectively unstable, with CO-rich material being efficiently dredged-up
from the outermost white dwarf layers.  This mechanism allows for metallicity enhancement of the envelope
through Kelvin-Helmholtz instabilities, at levels 
that agree with observations ($\sim 20-30$\%, by mass \cite{Geh98}).
Moreover, the simulations revealed that 
despite the differences found in the 
convective flow patterns in 1-D and 2-D models, the expansion and  progress of 
the  2-D burning front  towards the  outer  
envelope  quickly becomes almost spherically symmetric, even though the 
initial burning process was not. 

Another set of multidimensional simulations, aimed at verifying the general 
trends reported by Glasner et al., were performed by Kercek et al. \cite{KHT98} with a version of the Eulerian 
code {\it  PROMETHEUS}.
As in previous work, only a reduced computational domain 
(a box of about 1800 km$\times$1100 km), was used in the 2-D 
simulations. The results are characterized by 
somewhat less violent outbursts, longer runaways and lower peak temperatures than those found by Glasner et al.
These trends possibly resulted from
large differences in the convective  flow patterns: whereas Glasner et al. 
found that a few large convective eddies dominated the flow, Kercek et al. 
found that the early runaway was governed by small, very stable
eddies, which led to  more  limited dredge-up  and  mixing  episodes. 
Such discrepancies were even more striking in 3-D simulations\cite{KHT99},  
in which the limited dredge-up of core material translated into maximum envelope
velocities that were a  factor  $\sim  100$  smaller than  the  corresponding escape velocity;
presumably, no mass ejection resulted.  The controversy was carefully analyzed by Glasner et al. \cite{GLT05} who concluded that the early stages of the explosion, when the evolution is quasistatic, are extremely 
sensitive to the adopted outer
boundary conditions. 

Confirmation of the feasibility of the core-envelope mixing scenario was provided
by a set of independent 2-D simulations \cite{Cas10, Cas11a} performed
with the Eulerian code {\it FLASH}.  These models showed that 
 Kelvin-Helmholtz instabilities can naturally 
lead to self-enrichment of the accreted envelope with core material, at 
levels that agree with observations.

Two dimensional prescriptions for convection are, however, unrealistic \cite{AMY09}.
Indeed, the conservation of vorticity imposed by a 2-D geometry 
forces the small convective cells to merge into large eddies, 
with a size comparable to the pressure scale height of the envelope. 
In contrast, 
these eddies will become unstable and break up in 3-D (in fully developed turbulent convection), transferring their energy 
to progressively smaller scales\cite{Pop00, Sho07}. These smaller structures, vortices and filaments, 
will undergo a similar fate down to approximately the Kolmogorov scale. 
A pioneering 3-D simulation of mixing at the core-envelope interface during 
nova explosions \cite{Cas11b} has shown hints 
of the nature of the highly fragmented, chemically enriched and 
inhomogeneous nova shells observed in high-resolution spectra.  This, as predicted in the Kolmogorov theory of turbulence, has been 
interpreted as a relic of the hydrodynamic instabilities that develop 
during the initial ejection stage. 
Although such inhomogeneous patterns inferred from the ejecta have
usually been assumed to result from uncertainties in
the observational techniques, 
they may represent a real signature of 
the turbulence generated during the thermonuclear runaway. 

For X-ray bursts, only preliminary 2-D simulations of specific aspects of the explosions (such as
flame propagation \cite{Spi02} or the early convective stages preceding ignition \cite{Mal11}) have been conducted to date.  Attempts to overcome the difficulties associated with multidimensional modeling have included the filtering of acoustic waves.  This allows for larger time steps since, in this approximation, this quantity is now determined by
the fluid velocity rather than by the speed of sound. Several such ``low-Mach number" codes 
have been developed in recent years\cite{Spi02,Lin06,Mal11}.  It is not clear why more emphasis has been placed on the multidimensional modeling of novae rather than X-ray bursts.  We do note that some of the groups that performed pioneering multidimensional nova simulations (based in e.g., Arizona, Garching, and Chicago) eventually shifted towards the modeling of Type Ia supernovae. Only two groups (based in Jerusalem and Barcelona) are currently involved with multidimensional nova models, while only one (based in Stony Brook) is actively developing 2-D XRB simulations.

All of these multidimensional simulations are extremely time-consuming.  As a result, they rely only on very
simplified nuclear reaction networks (typically containing about a dozen species).  Furthermore, simulations have followed the evolution of a nova over only a very small fraction of the overall time associated with the event (e.g., $\sim 1000$ s near the peak temperature, to be
compared with the duration of the accretion stage for a nova outburst, $\sim 10^5$ yr).  Hence, no reliable nucleosynthesis predictions can be made using current multidimensional models. Finally, the use
of Eulerian frameworks do not allow one to follow the progress of the explosion once it reaches dynamic stages, as material would be artificially lost through the edges of the computational domain.  A 3-D implicit ALE hydrodynamic code would be best suited to overcome this limitation.

\section{When stellar explosions hate theorists: observational constraints}

\subsection{Absorption features in X-ray bursts}

The potential contribution of XRBs to Galactic abundances (e.g., of the light p-nuclei $^{92,94}$Mo and $^{96,98}$Ru) has been debated\cite{Sch98,Sch01,Wei06a,Baz08}.  Since only a few MeV per nucleon are released from thermonuclear fusion in XRBs (to be compared to $\approx200$ MeV per nucleon released through the accretion of matter onto a typical neutron star), ejection from a neutron star is energetically unlikely.  This has been confirmed by all recent hydrodynamic simulations\cite{Woo04,Fis08,Jos10}.  

Radiation-driven winds during photospheric radius expansion (PRE) may, however, lead to the
ejection of a tiny fraction (e.g., $\approx1$\%) of the envelope\cite{Kat83,Pac83,Nob94,Wei06a}.  PRE occurs in some bright type I XRBs when the luminosity reaches the Eddington limit and causes the expansion of the envelope.  This effect is usually first indicated through observation of a flat-topped light curve, with the bolometric luminosity being almost constant during the expansion of the envelope.  It can later be confirmed through time-resolved spectroscopy of the interval during which the luminosity is constant, through which the expansion and contraction of the photospheric radius (together with a decrease and increase of the effective temperature) may be deduced (see e.g., Ref.\cite{Str06} for examples of both PRE and non-PRE bursts from a single
source).  While these bursts allow one to estimate the mass and radius of the neutron star (and hence, constrain the neutron star equation of state\cite{Ste10,Dam90,Guv12a,Guv12b,Sal12}), it has not yet been analysed through detailed models whether such
winds may contain material synthesized during the burst.  Exacerbating the issue, in one-zone nucleosynthesis calculations the
final yields are often assumed to represent the composition of the entire envelope. In (multi-zone)
hydrodynamic simulations, however, the abundances of many species, including these p-nuclei, decrease by more than an order of magntiude relative to their values in the innermost layers because of limited convective transport\cite{Jos10}. Unfortunately it is material from these outermost layers that are most likely to be ejected by any radiation-driven winds. The predicted overproduction factors in these outer regions are several orders of magnitude smaller than those required to account for the origin of Galactic light p-nuclei\cite{Jos10,Baz08}, in contrast with the results reported from one-zone calculations\cite{Sch01}. As such, according to current models, XRBs are unlikely to be dominant contributors to the Galactic abundances of p-nuclei.

The detection of absorption lines in the neutron star atmosphere seems to be
the most direct and powerful technique for both probing the products of nucleosynthesis and for the determination of the neutron star equation of state (via measurement of the surface gravity).  The main observational difficulty lies in the short duration of the bursts: the detection of absorption features requires long exposure times to accumulate data with enough signal to noise.  Even for the most efficient and regular bursters (those with long and frequent bursts such as 
GS 1826-24, KS 1731-260 and IM 0836-425) with bursts of $20-40$ s in duration occurring 
every $2-3$ hours, only $\approx0.3$\% of the total observation time may be spent observing bursts.  Nonetheless, Cottam et al.\cite{gscottam2002} reported the first detection of gravitationally redshifted 
absorption lines in the X-ray spectrum of EXO 0748-676.  These measurements were possible because the source had been chosen for use in the calibration of the Reflection Grating Spectrometer
onboard the ESA X-ray observatory XMM-Newton\cite{XMM}, and as such, was observed for 335 ks.  Such long observations are not usually possible with X-ray observatories.  A total of 28 type I XRBs were recorded with the instrument, lasting a cumulative 3200 s.
That made possible the detection of several absorption features, among which the authors 
identified Fe and O lines that had been gravitationally redshifted by $z=0.35$.  Rauch et al.\cite{gsRauch} later suggested that the observed features correspond to different transitions than those suspected by Cottam et al., revealing the limitations inherent to models of neutron star atmospheres.  Unfortunately, a second, longer observation (almost 600 ks) of the same source with the same instrument failed to confirm the detection of the lines\cite{gscottam_failed}.  A long observation of the regular bursting source GS 1826-24 also failed to detect any 
absorption lines \cite{gskong}.  Recent, more encouraging results include the observation, albeit with limited spectral
resolution, of absorption edges from Fe-peak elements in photospheric radius ``superexpansion" bursts\cite{Zan10}.  Higher spectral resolution observations of these superexpansion bursts (e.g., with Chandra\cite{CHANDRA} or XMM-Newton\cite{XMM}, though an exposure time of $\approx500$ ks may be necessary) would be of great interest.

\subsection{Isotopic abundances from nova explosions}

Spectroscopic studies of classical nova explosions provide elemental abundances which can be used to constrain predictions of nucleosynthesis in nova models\cite{JH98,Sta00,Dow13,Kel13}.  Measurements of the relative abundances of different isotopes in nova ejecta could further improve model constraints;  these could be provided through measurements of presolar grains or through detections of $\gamma$-rays from the decay of radioisotopes produced during the explosion.

Classical novae are Galactic dust factories, as revealed by infrared and ultraviolet observations 
\cite{Geh98}.  The first to realize the importance of dust to constrain nova models
were Clayton and Hoyle \cite{CH76}, who pointed out a number of isotopic signatures that
should characterize grains condensed in nova ejecta (e.g.,  
large overproduction of $^{13,14}$C, $^{18}$O, $^{22}$Na, $^{26}$Al or $^{30}$Si relative to solar values). 
Since then, efforts to identify candidate grains from novae have focused mainly on
searching for low $^{20}$Ne/$^{22}$Ne ratios. This is because noble gases such
as Ne do not easily condense into grains; hence, excesses of $^{22}$Ne could be attributed to in situ decay of $^{22}$Na that had been trapped in the grain.  

In 2001 the first set of presolar SiC and graphite grains with isotopic compositions similar to nova model predictions were measured, after isolation from the Murchison and Acfer 094 meteorites \cite{Ama01,Ama02}.
These grains were characterized by low $^{12}$C/$^{13}$C and
$^{14}$N/$^{15}$N ratios, a high $^{30}$Si/$^{28}$Si ratio, and close-to-solar $^{29}$Si/$^{28}$Si values
 (high $^{26}$Al/$^{27}$Al and $^{22}$Ne/$^{20}$Ne ratios were also observed for some of these grains).   The composition of these grains, however, was only consistent with {\it diluted} abundances from model predictions. That is, mixing between nova ejecta predictions and more than ten times close-to-solar material prior to grain formation was required to match the grain composition.  Such mixing may result from the interaction
between the ejecta and the accretion disk, or even with the outer layers of the secondary star. 
Calculations to test such scenarios are currently in progress (Figueira et al., in preparation).
Later, concerns about the nova paternity of these grains were raised\cite{NH05} after the identification of three other SiC grains from
the Murchison meteorite with similar trends (in particular, low $^{12}$C/$^{13}$C and
$^{14}$N/$^{15}$N ratios) but additional features (such as non-solar Ti isotopic ratios).  As such, a supernova origin for these grains cannot be excluded\cite{NH05,JH07c}. The issue is far from being settled
since models suggest that rare, more violent nova outbursts in which the nuclear activity may extend beyond calcium
can be obtained in very-low metallicity systems \cite{Jos07first} or during mass-transfer episodes at
low rates \cite{Gla09}.

More recently, and as expected from an oxygen-rich (O$>$ C) environment, a few oxide grains 
whose composition can be qualitatively matched by nova models have also been identified \cite{Gyn10,Gyn11}.
Overall, only a handful of candidate grains from novae have been isolated.  As such, the implications derived from such analyses are not statistically sound. A larger number of grains of putative nova origin would be very valuable.

Isotopic abundances may also be provided through the detection of predicted $\gamma$-ray features from radioisotopes produced in novae \cite{CH74,Cla81,LC87,Gom98,Her08}.  A prompt $\gamma$-ray signature at and below 511 keV
(through electron-positron annihilation) is expected from $^{13}$N and $^{18}$F, while $^7$Be and $^{22}$Na, characterized by longer half-lives, would power line emission at 478 and 1275 keV, respectively.  Finally, because of the long lifetime of $^{26}$Al ($10^{6}$ years) it is only the cumulative emission of many novae superimposed on the emission from other $^{26}$Al production sites (rather than the emission from individual objects) that is expected to be observable.  The 511 keV line and the lower-energy continuum should exhibit the largest $\gamma$-ray fluxes of all of these characteristic nova signatures; unfortunately, these features would occur well before the visual
rise in luminosity, precluding the repointing of any space-borne observatory.  This restricts the search for this emission to archived data in the hope that a suitable instrument was serendipitously pointing towards the direction of the nova 
at the right time.  

The actual detection of $\gamma$-ray signatures around 1 MeV from novae has been elusive.  Efforts have aimed at identifying the $^{22}$Na line (1275 keV) using the COMPTEL instrument onboard the 
Compton Gamma-Ray Observatory, CGRO\cite{Iyu95,Iyu99}; and the 478 keV (from $^7$Be) and 1275 keV lines using the GRS instrument onboard the Solar Maximum Mission (SMM) satellite\cite{Lei88,Har91,Har96}. 
Other studies focused instead on the 511 keV line (and the lower energy continuum) arising from
electron-positron annihilation. Several instruments have been used to this end 
such as WIND/TGRS\cite{Har99,Har00} and CGRO/BATSE\cite{Her00}.  All of these investigations have only provided upper limits on the emission, all of which are fully compatible with theoretical
predictions. Note that the search for these $\gamma$-ray signatures from the decay of radioisotopes should not be confused with the recent FERMI-LAT observations\cite{Abd10}
of $\gamma$-rays emitted by four novae, at energies $E > 100$ MeV.  Such emission was likely produced in shocks within the ejecta
or in their interaction with the secondary star.  

Currently there are approved proposals aimed at monitoring nearby novae with
the INTEGRAL satellite\cite{Her08}; detection of $\gamma$-rays around 1 MeV with INTEGRAL would require a nova within roughly $\approx1$ kpc of Earth.  More sensitive $\gamma$-ray detectors with large collecting areas are needed to improve prospects for the unambiguous detection of $\gamma$-ray features
associated with the decay of radioactive species produced in novae.  Measurements of certain reaction rates (see Section \ref{sens}) would also help to improve predictions of the expected detectability distances of the emission.

\section{Improving stellar explosion models in the laboratory}
\label{sens}

An effective means to identify important nuclear interactions involves the examination of the impact on model
predictions from the systematic variation of each interaction in a network by its uncertainty.  Each rate may be varied
individually (requiring roughly as many model calculations as there are rates in the associated network) or a Monte Carlo method may be used to simultaneously vary all rates in the network (from which e.g., correlations between an enhancement factor applied to a rate X and
the corresponding change in the yield of a species Y may be deduced). Obviously, results from these types of sensitivity studies are most clearly interpreted when rates are varied by well-defined, temperature-dependent, experimentally-based uncertainties.  While this situation may hold for most reaction rates comprising networks coupled to standard models of classical nova explosions (as well as for, e.g., Big Bang nucleosynthesis), it certainly does not apply to networks needed for detailed nucleosynthesis predictions from type I X-ray bursts (especially for ignition conditions where an extended rp process is predicted).  Nonetheless, variation of theoretical rates in the network provides guidance to experimentalists as to where resources are best focused, as well as insight into the level of dependence of model predictions on the method used to determine theoretical rates.  Suitable
variation factors for theoretical rates may be adopted to account for possible discrepancies between predictions of rates from different codes.  For example, while these rates are often stated to be reliable, on average,
to a factor of $\approx2$, significantly larger deviations have been observed when comparing (i) statistical
model rates to experimental rates (up to a factor of $\approx100$ in some cases) and (ii) statistical model
rates for a common reaction determined with different codes (up to an order of magnitude)\cite{Par13rev}.  As well, different libraries of theoretical rates may be used to test the effects on model predictions of systematic differences (or improvements) in predicted rates.   

In this section we review studies that have either identified nuclear physics quantities of interest for predictions from models of classical novae and type I XRBs, or examined the impact of specific measurements on model predictions.  We do not discuss experimental or theoretical results that have not been demonstrated to significantly affect model predictions.  On this note, we encourage all relevant studies to test and report the impact of the obtained results in the framework of an astrophysical model whenever possible. 

\subsection {Sensitivity studies for classical nova explosions}

For standard models of classical nova explosions, detailed results from a comprehensive sensitivity study have only been reported in Iliadis et al.\cite{Ili02}, where post-processing of temperature-density-time trajectories from a single zone from each of five different hydrodynamic models was used along with the individual variation of 175 rates by factors of 2, 10 and 100 (up and down).  Reactions with significant impact on final nucleosynthesis predictions (e.g., modifying the yield of an isotope by at least a factor of two when varied by adopted uncertainties) were thereby identified.  [Note that the impact of a rate variation in one-zone studies is usually overestimated due to e.g., dilution effects from the composition of the outermost zones.  This effect may be modest ($\approx25\%$\cite{Sal11}) or considerable (many orders of magnitude\cite{Par03}) depending on the approaches and specific rates compared - see below.]  This study helped to motivate many experiments over the past decade to better constrain reactions involved in classical nova explosions; compare, e.g., the compilations of NACRE\cite{NACRE} and Iliadis et al.\cite{Ili01} with the later compilations of Iliadis et al.\cite{Ili10} and STARLIB\cite{Sal13}.  (For the appreciable impact of updated libraries of nuclear reaction rates on nucleosynthesis predictions from 1-D hydrodynamic nova models, see Starrfield et al.\cite{Sta98,Sta00,Sta09}.)  More limited studies have also been performed to examine e.g., the production of the radioisotopes $^{18}$F, $^{22}$Na and $^{26}$Al (Coc et al.\cite{Coc95}, using a semi-analytical treatment; Coc et al.\cite{Coc00}, Jos\'e et al.\cite{Jos99}, using a 1-D hydrodynamical model) and the production of certain elemental abundances that may help to assess the peak temperature or degree of mixing in a nova (Downen et al.\cite{Dow13} and Kelly et al.\cite{Kel13}, using post-processing of multiple zones from 1-D hydrodynamic models).  Partial results from comprehensive one-zone Monte Carlo sensitivity studies have also been reported\cite{Hix03,Smi02}, where uncertainties were assumed as $\approx$50\% for rates whose measurement would require radioactive beams, a factor of two for rates calculated through statistical models, and $\approx$20\% for all other non-weak interaction rates.  

We review the impact of the most influential reactions identified in these sensitivity studies below.  (Note that nuclear masses and weak interaction rates seem to be sufficiently well known for current nova models.)  Often, though not always, an article on a relevant experiment explores the impact of the outstanding uncertainty in a rate using some model.  We also include below some of these particular results with significant impact on nucleosynthesis predictions from models.

\subsubsection*{Thermonuclear reaction rates with demonstrated impact on nova model predictions:}
\begin{itemize}

\item $^{17}$O($p,\gamma$)$^{18}$F and $^{17}$O($p,\alpha$)$^{14}$N 

Using a 1-D hydrodynamic model, Coc et al.\cite{Coc00} found that their calculated $^{18}$F abundance varied by a factor of $\approx10$ when comparing results using the NACRE\cite{NACRE} low and high $^{17}$O+p rates (at 0.2 GK, the ratio of the high to low rate is about two orders of magnitude for both the ($p,\gamma$) and ($p,\alpha$) rates).  Later, when the ($p,\gamma$) and ($p,\alpha$) rates were individually varied by the adopted uncertainties (a maximum of a factor of ten over T$_{peak}$ = $0.1 - 0.4$ GK) in the post-processing studies of Iliadis et al.\cite{Ili02}, the final calculated abundances of $^{17}$O and $^{18}$F varied by at least a factor of two.  In a one-zone post-processing study conducted with a Monte Carlo approach to rate variations, both of these are given in a prioritized list of reactions that affect the production of the radioisotope $^{18}$F\cite{Smi02,Hix03}.  Later, the impact of the uncertainties in these rates as determined by Fox et al.\cite{Fox04} ($\approx30\%$ and a factor of $\approx2.5$ for the ($p,\gamma$) and ($p,\alpha$) rates, respectively, at T = 0.2 GK) was examined using a 1-D hydrodynamic model.  Through comparing yields calculated in a model with the previous rate uncertainties (a few orders of magnitude and an order of magnitude for the ($p,\gamma$) and ($p,\alpha$) rates, respectively, at T = 0.2 GK) to yields from a model with their deduced uncertainties for these rates, Fox et al. found, e.g., that the range of final abundances of $^{18}$O, $^{18}$F, and $^{19}$F was reduced from $\approx1-2$ orders of magnitude to less than a factor of three.  Chafa et al.\cite{Chaf05} used a similar 1-D hydrodynamic model to examine the effect of their enhanced $^{17}$O($p,\alpha$) rate relative to that of Fox et al.  (The ratio of their $^{17}$O($p,\alpha$) rate to the $^{17}$O($p,\gamma$) rate was $\approx100$ times greater than that from Fox et al. or from NACRE\cite{NACRE}, at 0.2 GK.)   They found that with their new $^{17}$O+p rates, the final abundances of $^{17}$O and $^{18}$F were reduced by factors of $2 - 3$ relative to those from Fox et al., and by factors of 1.4 and 8, respectively, relative to those using the NACRE rates.  Moazen et al.\cite{Moa07} tested the impact of similar rates to those of Chafa et al. (in agreement to better that 5\% over nova temperatures) using a multi-zone post-processing approach\cite{Par03} (where final abundances are determined by neglecting any mixing and simply summing the contributions of each zone weighted by the total mass of the zone).  For a similar mass white dwarf (1.15 M$_{\odot}$) to that used in the hydrodynamic tests mentioned above, they find reductions by a factor of 10 in the abundance of $^{18}$F relative to the abundance of $^{18}$F determined using the NACRE $^{17}$O+p rates.

\item $^{17}$F($p,\gamma$)$^{18}$Ne

When individually varied by the adopted uncertainty (a maximum of a factor of ten over T$_{peak}$ = $0.1 - 0.4$ GK) in the post-processing studies of Iliadis et al.\cite{Ili02}, the final calculated abundances of $^{17}$O and $^{18}$F varied by at least a factor of two.  In a one-zone post-processing study conducted with a Monte Carlo approach to rate variations, this reaction is given in a prioritized list of reactions that affect the production of the radioisotope $^{18}$F\cite{Smi02,Hix03}.  In Parete-Koon et al.\cite{Par03}, a multi-zone post-processing method found that abundances of $^{15}$N, $^{17}$O, $^{18}$O, $^{18}$F, and $^{19}$F vary by factors of $\approx2 -4$ (depending on the mass of the white dwarf considered) when using rates that differ by as much as a factor of $\approx30$ at 0.4 GK.     
The multi-zone post-processing study of Chipps et al.\cite{Chi09} found that the final abundances of $^{17}$O and $^{18}$F vary by less than a factor of two when rates which differ by a factor of $\approx14$ at 0.2 GK were used.  (Note that implications of uncertainties in the important direct capture component of this rate were not considered in this study.)

\item $^{18}$F($p,\gamma$)$^{19}$Ne and $^{18}$F($p,\alpha$)$^{15}$O

Using a 1-D hydrodynamic model, Coc et al.\cite{Coc00} found that the calculated $^{18}$F abundance varied by a factor of $\approx300$ when comparing results using their low and high $^{18}$F+p rates (at 0.2 GK, the ratio of the high to low rate was about one and two orders of magnitude for the ($p,\gamma$) and ($p,\alpha$) rates, respectively).  When individually varied by the adopted uncertainties (a maximum of a factor of 15 and a factor of 30 for the ($p,\gamma$) and ($p,\alpha$) rates, respectively, over T$_{peak}$ = $0.1 - 0.4$ GK) in the post-processing studies of Iliadis et al.\cite{Ili02}, the final calculated abundances of $^{16}$O, $^{17}$O and $^{18}$F varied by at least a factor of two due to variations in the ($p,\alpha$) rate.  In a one-zone post-processing study conducted with a Monte Carlo approach to rate variations, the $^{18}$F($p,\alpha$) reaction is given in a prioritized list of reactions that affect the production of the radioisotope $^{18}$F\cite{Smi02,Hix03}.  Using a multi-zone post-processing approach\cite{Par03}, Bardayan et al.\cite{Bar02} and Kozub et al.\cite{Koz05} reported that roughly twice as much $^{18}$F was produced when using a ($p,\alpha$) rate $\approx1.5 - 2$ times lower and $\approx5$ times lower, respectively, than that of Coc et al.\cite{Coc00} at nova temperatures.  Chae et al.\cite{Chae06}, also using a multi-zone approach, found that the amount of $^{18}$F produced varies by roughly a factor of two when ($p,\alpha$) rates differing by roughly a factor of $\approx10$ are used (see also Adekola et al.\cite{Ade11}).  Laird et al.\cite{Lai13}, using a 1-D hydrodynamic model, found that $^{18}$F production varies by roughly a factor of two when two ($p,\alpha$) rates differing by roughly a factor of two (at 0.2 GK) are used.

\item $^{21}$Na($p,\gamma$)$^{22}$Mg

Coc et al.\cite{Coc95}, using a semi-analytical approach (a ``compromise" between hydrodynamic and one-zone calculations), found that the final abundance of the radioisotope $^{22}$Na increased by a factor of $\approx1.5$ when their $^{21}$Na($p,\gamma$) rate was reduced by a factor of 10, and increased by a factor of $\approx1.5 - 3$ (depending on the white dwarf mass) when their $^{21}$Na($p,\gamma$) rate was reduced by a factor of 100.  Using a 1-D hydrodynamic model, Jos\'e et al.\cite{Jos99} also found that $^{22}$Na production increased by factors of $\approx1.5 - 3$ (depending on the white dwarf mass) when their rate was reduced by a factor of 100.  When individually varied by the adopted uncertainty (a factor of 100) in the post-processing studies of Iliadis et al.\cite{Ili02}, the final calculated abundances of $^{21}$Ne, $^{22}$Na and $^{22}$Ne varied by at least a factor of two.  In a one-zone post-processing study conducted with a Monte Carlo approach to rate variations, the $^{21}$Na($p,\gamma$) reaction is given in a prioritized list of reactions that affect the production of $^{22}$Na\cite{Smi02,Hix03}.  Using a 1-D hydrodynamic model, Bishop et al.\cite{Bis03} and Davids et al.\cite{Dav03} found that $^{22}$Na production decreased by about 20\% when results using two rates differing by a factor of $\approx5$ (at 0.2 GK) were compared.

\item $^{22}$Na($p,\gamma$)$^{23}$Mg

Coc et al., using a semi-analytical approach\cite{Coc95}, compared the yield of $^{22}$Na obtained using different $^{22}$Na($p,\gamma$) rates.  Reduced rates by a factor of $\approx10$ or $\approx50$ at 0.2 GK  increased the $^{22}$Na yield by a factor of $\approx10$ or $\approx40$, respectively, in their models.  Using a 1-D hydrodynamic model, Jos\'e et al.\cite{Jos99} found that $^{22}$Na production increased by a factor of $\approx3$ when comparing results using rates that differed by more than one order of magnitude for T $>$ 0.1 GK.  In a one-zone post-processing study conducted with a Monte Carlo approach to rate variations, the $^{22}$Na($p,\gamma$) reaction is first in a prioritized list of reactions that affect the production of $^{22}$Na\cite{Smi02,Hix03}.  Jenkins et al.\cite{Jen04} and Sallaska et al.\cite{Sal11} compared yields from hydrodynamic models using two $^{22}$Na($p,\gamma$) rates differing by a factor of $\approx10$ and a factor of $\approx3$ (both at 0.2 GK), respectively; they found that the $^{22}$Na yield varied by a factor of $\approx3$ and $\approx2$, respectively.

\item $^{23}$Na($p,\gamma$)$^{24}$Mg and $^{23}$Na($p,\alpha$)$^{20}$Ne

Using a 1-D hydrodynamic model, Jos\'e et al.\cite{Jos99} found that when comparing results obtained using $^{23}$Na($p,\gamma$) rates differing by factors of $3-1000$ over $0.1 - 0.2$ GK, yields of $^{22}$Na, $^{25}$Mg, $^{26}$Al and $^{27}$Al varied by $\approx30\%$ and the yield of $^{24}$Mg varied by a factor of $\approx3$.  When individually varied by the adopted uncertainties (a maximum of a factor of 10 and 1.4 for the ($p,\gamma$) and ($p,\alpha$) rates, respectively, over T$_{peak}$ = $0.1 - 0.4$ GK) in the post-processing studies of Iliadis et al.\cite{Ili02}, the variations in the $^{23}$Na($p,\gamma$) rate affected the final calculated abundances of many species between $^{20}$Ne and $^{27}$Al by at least a factor of two.   In a one-zone post-processing study conducted with a Monte Carlo approach to rate variations, the $^{23}$Na($p,\gamma$) and $^{23}$Na($p,\alpha$) reactions are both given in a prioritized list of reactions that affect the production of both $^{22}$Na and $^{26}$Al\cite{Smi02,Hix03}.  Rowland et al.\cite{Row04} used 1-D hydrodynamic models to test the effect of improved uncertainties for the $^{23}$Na+p rates.  For example, at 0.3 GK, the uncertainty in their ($p,\alpha$)/($p,\gamma$) rate ratio was $\approx30\%$, while that from NACRE\cite{NACRE} was a factor of $\approx3$ (with similar central value).  This led to improved constraints on the production of isotopes between $^{22}$Na and $^{29}$Si; e.g., the uncertainty in the production of $^{26}$Al was reduced from a factor of $\approx3$ to $\approx20\%$. 

\item $^{23}$Mg($p,\gamma$)$^{24}$Al

When individually varied by the adopted uncertainty (a factor of 100) in the post-processing studies of Iliadis et al.\cite{Ili02}, the $^{23}$Mg($p,\gamma$) rate affected the final calculated abundances of Ne, Na and Mg isotopes by at least a factor of two.   In a one-zone post-processing study conducted with a Monte Carlo approach to rate variations, the $^{23}$Mg($p,\gamma$) reaction is first in a prioritized list of reactions that affect the production of $^{26}$Al\cite{Smi02,Hix03}.  

\item $^{25}$Al($p,\gamma$)$^{26}$Si

Coc et al., using a semi-analytical approach\cite{Coc95}, compared yields of $^{26}$Al determined using several different $^{25}$Al($p,\gamma$)$^{26}$Si rates.  For two extreme rates (differing by five orders of magnitude at 0.2 GK), the $^{26}$Al yields varied by a factor of $\approx5$ in their models.  Using a 1-D hydrodynamic model, Jos\'e et al.\cite{Jos99} found that when comparing results using the same extreme rates of Coc et al.\cite{Coc95}, the $^{26}$Al yield varied by a factor of $\approx2$.  In a one-zone post-processing study conducted with a Monte Carlo approach to rate variations, the $^{25}$Al($p,\gamma$) reaction is given in a prioritized list of reactions that affect the production of $^{26}$Al\cite{Smi02,Hix03}.  Bardayan et al.\cite{Bar06} used a multi-zone post-processing approach\cite{Par03} and found that a rate uncertainty of roughly two orders of magnitude (at 0.2 GK) resulted in variation of the $^{26}$Al yield by a factor of $\approx3$.  Using a 1-D hydrodynamic model, Parikh and Jos\'e\cite{Par13} found no significant change to any predicted yields when comparing results using two $^{25}$Al($p,\gamma$) rates that differed by factors of $6 - 50$ over T = $0.05 - 0.2$ GK.  Very recently, Bennett et al.\cite{Ben13} report no significant change in predicted $^{26}$Al yields (as determined with a 1-D hydrodynamic model) when their uncertainty for the $^{25}$Al($p,\gamma$)$^{26}$Si rate is employed.    

\item $^{26g,m}$Al($p,\gamma$)$^{27}$Si

Coc et al., using a semi-analytical approach\cite{Coc95}, compared yields of $^{26}$Al determined using several different $^{26g}$Al($p,\gamma$)$^{27}$Si rates.  For two extreme rates (varying by many orders of magnitude below 0.1 GK, but nearly identical at 0.2 GK), the $^{26}$Al yields varied by a factor of only $\approx10\%$ in their models.  They mention, however, that the $^{26}$Al yield may vary by a factor of $2-3$ if the strength of a resonance at 188 keV in the $^{26g}$Al($p,\gamma$) reaction is (arbitrarily) scaled by a factor of three (see below).  Using a 1-D hydrodynamic model, Jos\'e et al.\cite{Jos99} found a reduction of $^{26}$Mg (by a factor of $\approx2$) and no significant effect on the $^{26}$Al yield from increasing their $^{26m}$Al($p,\gamma$) rate by a factor of 100.  They also agreed with the results of Coc et al.\cite{Coc95} on the dependence of the $^{26}$Al yield on the strength of the 188 keV resonance in the $^{26g}$Al($p,\gamma$) reaction.  When individually varied by the adopted uncertainties (changes by a factor of 10 (up) and 0.8 (down) for the $^{26g}$Al($p,\gamma$) rate, and a factor of 100 (up and down) for the $^{26m}$Al($p,\gamma$) rate) in the post-processing studies of Iliadis et al.\cite{Ili02}, the $^{26g}$Al($p,\gamma$) and $^{26m}$Al($p,\gamma$) rates affected the final calculated abundances of $^{26}$Al and $^{26}$Mg, respectively, by at least a factor of two.  In a one-zone post-processing study conducted with a Monte Carlo approach to rate variations, the general $^{26}$Al($p,\gamma$) reaction is given in a prioritized list of reactions that affect the production of $^{26}$Al\cite{Smi02,Hix03}.   Ruiz et al.\cite{Rui06} remeasured the strength of the 188 keV resonance in $^{26g}$Al($p,\gamma$) and found it to be $\approx2/3$ of the previously assumed value\cite{Coc95,Jos99}.  Using a 1-D hydrodynamic model, they compared results using two rates that differed by $\approx20\%$ over nova temperatures and found a variation in the $^{26}$Al yield of about $20\%$.

\item $^{30}$P($p,\gamma$)$^{31}$S

When individually varied by the adopted uncertainty (a factor of 100) in the post-processing studies of Iliadis et al.\cite{Ili02}, the $^{30}$P($p,\gamma$) rate affected the final calculated abundances of a number of species between $^{30}$Si and $^{38}$Ar by at least a factor of two.  Using a 1-D hydrodynamic model, Jos\'e et al.\cite{Jos01} found that reduction of their $^{30}$P($p,\gamma$)$^{31}$S rate by a factor of 100 changed
abundances within A = $30-38$ by factors of $2-10$.  Ma et al.\cite{Ma07}, using a multi-zone post-processing approach\cite{Par03}, found that production of species between Si and Ca varied by as much as 40\% when two rates that differed by as much as a factor of $\approx10$ (around 0.2 GK) were used.  Parikh et al.\cite{Par11}, using a 1-D hydrodynamic model, found that rates that differed by as much as a factor of $\approx20$ (at 0.3 GK) led to a factor of up to $\approx3$ variation in yields of Si--Ar isotopes.
Downen et al.\cite{Dow13} and Kelly et al.\cite{Kel13} used a multi-zone post-processing approach together with individual rate variations and simultaneous variations (through a Monte Carlo method), respectively.  For their proposed ``mixing meters" and ``nova thermometers", they found that rate uncertainties have the largest impact on the predicted elemental abundance ratios of Si/H, O/S, S/Al, O/P and P/Al, with the dominant rate uncertainty being that of the $^{30}$P($p,\gamma$) reaction.

\item $^{33}$S($p,\gamma$)$^{34}$Cl

When individually varied by the adopted uncertainty (a factor of 100) in the post-processing studies of Iliadis et al.\cite{Ili02}, the $^{33}$S($p,\gamma$) rate affected the final calculated abundances of $^{33}$S, $^{34}$S, $^{35}$Cl and $^{36}$Ar by at least a factor of two.  Fallis et al.\cite{Fal13}, using a 1-D hydrodynamic model, determined the impact on nova nucleosynthesis of several different $^{33}$S($p,\gamma$) rates.  For example, rates differing by a factor of $\approx3$ at 0.3 GK modified the yield of $^{33}$S by a factor of $\approx4$.

\end{itemize}

Other important reaction-rates identified in the post-processing studies of Iliadis et al.\cite{Ili02} include $^{22}$Ne($p,\gamma$)$^{23}$Na, $^{26}$Mg($p,\gamma$)$^{27}$Al, $^{29}$Si($p,\gamma$)$^{30}$P, $^{33}$Cl($p,\gamma$)$^{34}$Ar, $^{34}$S($p,\gamma$)$^{35}$Cl, $^{34}$Cl($p,\gamma$)$^{35}$Ar, $^{37}$Ar($p,\gamma$)$^{38}$K, and $^{38}$K($p,\gamma$)$^{39}$Ca; other rates affecting the production of radioisotopes, as identified by Smith et al.\cite{Smi02} and Hix et al.\cite{Hix03} include $^{16}$O($p,\gamma$)$^{17}$F, $^{20}$Ne($p,\gamma$)$^{21}$Na, and $^{25}$Mg($p,\gamma$)$^{26}$Al.  Hydrodynamic tests using current uncertainties for these rates (and others mentioned above, such as $^{23}$Mg($p,\gamma$)) are encouraged.  Finally, we note that rates of the $^{14}$O($\alpha,p$) and $^{15}$O($a,\gamma$) reactions have no significant impact on nova nucleosynthesis in standard models\cite{Ili02,Dav03b,Smi02, Bla03}.  Some nova models that explore quite extreme, rare scenarios (low accretion rates\cite{Gla09} or accretion of extremely metal-poor material\cite{Jos07first}) have been published in recent years.  It may be interesting to explore the role of nuclear uncertainties in these models, for completeness.

In summary, most of the nuclear physics uncertainties for models of novae have been identified and addressed by experiments. In particular, significant recent progress has been made towards better determining the rates of the $^{18}$F(p,$\alpha$)$^{15}$O\cite{Bee11, Ade11a, Ade11b, Ade12, Lai13,Gul13}, $^{25}$Al(p,$\gamma$)$^{26}$Si\cite{Pep09,Chi10,Mat10,Che12,Par13,Ben13} and $^{30}$P(p,$\gamma$)$^{31}$S\cite{Wre09,Par11,Doh12,Irv13} reactions, often stated as dominant contributors to remaining uncertainties in nova nucleosynthesis\cite{Jos07rev,Jos11rev}.  Nevertheless, a comprehensive sensitivity study using a hydrodynamic model with current rate uncertainties\cite{Cyb10,Ili10,Sal13} should be performed to end the debate of whether studies based solely on post-processing treatments are fully reliable.

\subsection {Sensitivity studies for type I X-ray bursts}

For models of type I XRBs with parameters (e.g., accretion rate, composition of accreted material) chosen to favour nucleosynthesis that eventually proceeds via the rp process, detailed results from a comprehensive sensitivity study have only been reported in Parikh et al.\cite{Par08,Par09}.  Post-processing of temperature-density-time trajectories from single zones (sampling parameter space in peak temperature, burst duration and composition of accreted material) was used with both the individual variation of $\approx3500$ rates and the simultaneous variation of all rates using a Monte Carlo method (see also Roberts et al.\cite{Rob06} for preliminary results from a post-processing sensitivity study using a Monte Carlo method for rate variations).  Reactions influencing the predicted nucleosynthesis, as identified using the two methods, were generally in excellent agreement\cite{Par08}.  (We note that studies have shown that, at least for one-zone models, there appears to be a strong correlation between the sensitivity of the burst ashes and the sensitivity of the light curve to individual rates\cite{AmtPhD,Smi08}.)  Furthermore, reaction $Q$-values (and hence, nuclear masses) with uncertainties sufficiently large to significantly impact predicted yields were identified (see also Brown et al.\cite{Bro02} and Fleckenstein\cite{FleDip} for important masses identified through one-zone models).   Knowledge of these masses is critical to understanding the (p,$\gamma$)--($\gamma$,p) reaction rate equilibria that develop at waiting points along the rp process and the subsequent evolution of abundances beyond these nuclei\cite{Sch98, Bro02, Sch06}.  Weak interaction rates as determined in the laboratory were also varied in this study, by both experimental uncertainties (which had no impact on final yields\cite{Par08}) and larger factors.  Although this approach can also probe the influence of a particular nuclear physics uncertainty on the predicted nuclear energy generation rate E$_{gen}$ during a burst, it is usually not sufficient to examine in detail the impact on predictions of the chief observable, the XRB light curve.  This is because these one-zone studies neglect crucial hydrodynamic effects such as convection and the finite diffusion time for energy to escape from the accreted envelope.  Furthermore, the post-processing approach is not self-consistent; that is, the thermodynamic history employed is independent of changes in the nuclear energy generation rate due to a rate variation.  

Further progress has been made through (i) calibrating a one-zone model to a 1-D hydrodynamic model by adjusting ignition
conditions until the light curve and final ashes agree as much as possible, (ii) identifying reaction rates (through the individual variation approach) to which the calibrated one-zone model shows the greatest sensitivity, followed by (iii) considering these rate variations in a 1-D hydrodynamic model to assess the impact on light curves.  Preliminary and partial results from such studies have been reported\cite{AmtPhD, Amt06, Smi08}.  The impact of different rate libraries\cite{Sch98, Koi99, Cyb10}, different sets of proton separation energies (calculated using different compilations of masses)\cite{Bro02,Clem03, Kan12, FleDip}, different network sizes\cite{Wal81, Koi04}, and variations in groups of weak interaction rates\cite{Woo04} has also been explored.  

We summarize the impact of the most influential rates and masses from these sensitivity studies below.  We also include below results from limited investigations of the impact of individual reaction rate uncertainties (e.g., from explorations of the impact of a new rate calculated following an experiment) on predictions from 1-D hydrodynamic or one-zone models coupled to large networks.  

\subsubsection*{Nuclear physics quantities with demonstrated impact on XRB model predictions:}

\paragraph{Thermonuclear reaction rates}
\begin{itemize}

\item$^{14}$O($\alpha,p$)$^{17}$F

When individually varied by the adopted uncertainty (a factor of 10) in the post-processing studies of Parikh et al.\cite{Par08}, the $^{14}$O($\alpha,p$) rate led to significant variations (by greater that 5\%) in the nuclear energy generation rate E$_{gen}$ during a burst.   Hu et al.\cite{He13a}, also using a one-zone post-processing model, also found significant changes to the profile of E$_{gen}$ vs time when comparing two rates that differed by factors of $2-36$ over $0.1-2$ GK.  The use of two other rates that differed by a factor of $\approx5$ at 0.35 GK and less than 10\% at 1 GK led to essentially identical E$_{gen}$ profiles in that study.

\item$^{15}$O($\alpha,\gamma$)$^{19}$Ne

In a post-processing study conducted with a Monte Carlo approach to rate variations\cite{Rob06}, the $^{15}$O($\alpha,\gamma$) reaction rate is stated to affect the nuclear energy generation rate during early stages of the burst.  When individually varied by a factor of 3 or 10 in the post-processing studies of Parikh et al.\cite{Par08}, the $^{15}$O($\alpha,\gamma$) rate led to significant variations (by greater that 5\%) in the nuclear energy generation rate E$_{gen}$ during a burst.  Fisker et al.\cite{Fis06,Fis07} and Davids et al.\cite{Dav11}, using different 1-D hydrodynamic models, varied the $^{15}$O($\alpha,\gamma$) rate within uncertainties to determine its impact on light curves. When Fisker et al. used a previous ``lower limit" for the rate\cite{Fis06}, their model revealed non-bursting behaviour - more precisely, a slowly oscillating luminosity with a period of about
four hours.  The use of a larger rate (by a factor of $\approx10$ at 0.5 GK) produced bursts in their model, with L$_{peak}$ about two orders of magnitude greater than luminosities they found with the lower rate\cite{Fis07}.  On the other hand, when Davids et al. used a very similar rate to the above-mentioned ``lower limit"\cite{Fis06} (within a model with similar accretion rate as well) their simulation not only produced bursts (in apparent contradiction with the results from Fisker et al.\cite{Fis06}), but the
brightest bursts; namely, a \emph{larger} $^{15}$O($\alpha,\gamma$) rate resulted in bursts with $\approx50\%$ \emph{lower} L$_{peak}$ values in their model\cite{Dav11}.

\item$^{18}$Ne($\alpha,p$)$^{21}$Na

When individually varied by a factor of 3 or 10 in the post-processing studies of Parikh et al.\cite{Par08}, the $^{18}$Ne($\alpha,p$) rate led to significant variations in the nuclear energy generation rate E$_{gen}$ during a burst (by greater that 5\%, see also Groombridge et al.\cite{Gro02}); variation of the rate by a factor of 10 also significantly changed the final yields of at least three species by at least a factor of two.  When varied by a factor of 100 in a one-zone model ``calibrated" to reproduce results from a 1-D hydrodynamic code\cite{AmtPhD}, the rate of this reaction led to significant changes in the predicted light curve.   When later varied by a factor of 30 in a 1-D hydrodynamic model\cite{AmtPhD}, only minor changes to the predicted light curve were observed.  Matic et al.\cite{Mat09} and He et al.\cite{He13b} examined the impact of this rate using a 1-D hydrodynamic model and one-zone post-processing calculations, respectively.  At 0.5 GK, the reaction rates determined in these two studies agree to $\approx10\%$ and are larger than that of Gorres et al.\cite{Gor95} by a factor of $\approx100$.  Matic et al. find that their larger rate (relative to that of Gorres et al.) gives a slightly lower predicted L$_{peak}$\cite{Mat09}; He at al., on the other hand, find the opposite trend in predicted E$_{gen}$ during the burst\cite{He13b}.

\item$^{23}$Al($p,\gamma$)$^{24}$Si

In a post-processing study conducted with a Monte Carlo approach to rate variations\cite{Rob06}, the $^{23}$Al($p,\gamma$) reaction rate is stated to strongly influence E$_{gen}$ and the nucleosynthesis during the burst.  When individually varied by a factor of 10 in the post-processing studies of Parikh et al.\cite{Par08}, the $^{23}$Al($p,\gamma$) rate led to significant variations (by greater that 5\%) in the nuclear energy generation rate E$_{gen}$ during a burst.   When varied by a factor of 100 in a one-zone model ``calibrated" to reproduce results from a 1-D hydrodynamic code and later by a temperature-dependent uncertainty in a 1-D hydrodynamic model\cite{AmtPhD, Smi08}, the rate of this reaction led to significant changes in the predicted light curve. 

\item$^{30}$S($\alpha,p$)$^{33}$Cl

Fisker et al.\cite{Fis04} multiplied the rate of the $^{30}$S($\alpha,p$) reaction by a factor of 100, finding minor effects relevant to a double peak structure in their predicted light curves.  When individually varied by a factor of 10 in the post-processing studies of Parikh et al.\cite{Par08}, the $^{30}$S($\alpha,p$) rate led to significant variations in both the nuclear energy generation rate E$_{gen}$ during a burst (by greater that 5\%) and the final yields (affecting the yields of at least three species by at least a factor of two).  When varied by a factor of 100 in a one-zone model ``calibrated" to reproduce results from a 1-D hydrodynamic code\cite{AmtPhD}, the rate of this reaction led to significant changes in the predicted light curve.   When later varied by a factor of 10 in a 1-D hydrodynamic model\cite{AmtPhD}, only minor changes to the predicted light curve were observed. 

\item$^{34}$Ar($\alpha,p$)$^{37}$K

Fisker et al.\cite{Fis04} multiplied the rate of the $^{34}$Ar($\alpha,p$) reaction by a factor of 100, finding minor effects relevant to a double peak structure in their predicted light curves.  When varied by a factor of 100 in a one-zone model ``calibrated" to reproduce results from a 1-D hydrodynamic code\cite{AmtPhD}, the rate of this reaction led to significant changes in the predicted light curve.   When later varied by a factor of 10 in a 1-D hydrodynamic model\cite{AmtPhD}, only minor changes to the predicted light curve were observed. 

\item$^{59}$Cu($p,\gamma$)$^{60}$Zn

When individually varied by a factor of 10 in the post-processing studies of Parikh et al.\cite{Par08}, the $^{59}$Cu($p,\gamma$) rate led to significant variations in both the nuclear energy generation rate E$_{gen}$ during a burst (by greater that 5\%) and the final yields (affecting the yields of at least three species by at least a factor of two).  When varied by a factor of 100 in a one-zone model ``calibrated" to reproduce results from a 1-D hydrodynamic code and later by the same factor in a 1-D hydrodynamic model\cite{AmtPhD, Smi08}, the rate of this reaction led to significant changes in the predicted light curve. 

\end{itemize}

Other important reaction-rates mentioned in both Parikh et al.\cite{Par08} and Amthor\cite{AmtPhD} include: $^{22}$Mg($\alpha,p$), $^{25}$Si($\alpha,p$), $^{29}$S($\alpha,p$), $^{31}$Cl($p,\gamma$), $^{57}$Cu($p,\gamma$), $^{61}$Ga($p,\gamma$) and $^{65}$As($p,\gamma$).  Of these, all but $^{65}$As($p,\gamma$) led to significant changes in the predicted light curve of the calibrated one-zone model when varied by a factor of 100; only the $^{22}$Mg($\alpha,p$) rate, however, when varied by a factor of 10, had a significant effect in follow-up studies with a 1-D hydrodynamic model\cite{AmtPhD}.  As well, of these additional reactions, only the $^{31}$Cl($p,\gamma$) and $^{65}$As($p,\gamma$) reactions had a significant effect on predictions of both final yields and E$_{gen}$ in the post-processing sensitivity study (when varied by a factor of 10)\cite{Par08}.  We also note results in Thielemann et al.\cite{Thi01} where, using a 1-D hydrodynamic model, dramatic differences in light curves calculated with two
networks identical except for four rates (proton-capture on $^{27}$Si, $^{31}$S, $^{35}$Ar and $^{38}$Ca) were reported.\\

\paragraph{Nuclear masses}

To constrain models of type I XRBs, nuclear masses are desired to a precision of better than $\approx50$ keV\cite{Sch06,Par09}.  Through studies with a one-zone model, Brown et al.\cite{Bro02} identified species along the path of the rp process for which a more precise mass would better constrain their calculations: $^{61}$Ga, $^{62}$Ge, $^{64}$Ge, $^{65}$As, $^{66}$Se, $^{68}$Se, $^{69}$Br, $^{70}$Kr, $^{72}$Kr, $^{73}$Rb, and $^{74}$Sr.  Later, following post-processing sensitivity studies in which mass measurements needed to better constrain reaction $Q$-values were examined, Parikh et al.\cite{Par09} encouraged measurements of the masses of $^{26}$P, $^{27}$S, $^{43}$V, $^{46}$Mn, $^{47}$Mn, $^{51}$Co, $^{56}$Cu, $^{62}$Ge, $^{65}$As, $^{66}$Se, $^{69}$Br, $^{70}$Kr, $^{84}$Nb, $^{85}$Mo, $^{86}$Tc, $^{87}$Tc, $^{89}$Ru, $^{90}$Rh, $^{96}$Ag, $^{97}$Cd, $^{99}$In, $^{103}$Sn, $^{106}$Sb, and $^{107}$Sb, and more precise measurements of the masses of $^{31}$Cl, $^{45}$Cr, and $^{61}$Ga, $^{71}$Br, $^{83}$Nb, and $^{86}$Mo.  Fleckenstein\cite{FleDip}, using a one-zone XRB model, assessed the role of uncertainties in masses of species in the region A = $80 - 105$, and identified the most influential masses as $^{94}$Ag, $^{93}$Pd, $^{91}$Rh, $^{94}$Pd, $^{80}$Zr, $^{95}$Ag, $^{90}$Ru, $^{99}$In, $^{98}$Cd, $^{91}$Ru, $^{103}$Sn, and $^{100}$In.  

Of the above masses, according to the 2012 Atomic Mass Evaluation\cite{AME12}, experimental determinations of the masses of $^{26}$P, $^{27}$S, $^{46}$Mn, $^{56}$Cu, $^{62}$Ge, $^{66}$Se, $^{70}$Kr, $^{73}$Rb, $^{74}$Sr, $^{84}$Nb, $^{86}$Tc, $^{89}$Ru, $^{90}$Rh, $^{97}$Cd, $^{99}$In, $^{94}$Ag, $^{93}$Pd, $^{91}$Rh, and $^{95}$Ag are still needed. In addition, the experimentally-known masses of $^{31}$Cl, $^{61}$Ga, $^{65}$As, $^{80}$Zr, $^{83}$Nb, $^{96}$Ag, $^{100}$In and $^{103}$Sn may be needed to better precision.  We note that the impact on predictions from XRB models coupled to large networks has been examined for recent measurements of particular species (most mentioned above), including: $^{45}$Cr\cite{Yan13},  $^{65}$As\cite{Tu11}, $^{68}$Se\cite{Sav09}, $^{69}$Br\cite{Rog11}, $^{87}$Tc\cite{Hae11}, $^{99}$Cd\cite{Bre09}, $^{105}$Sb\cite{Maz07} and $^{106}$Sb\cite{Elo09}.

Recent experimental studies such as measurements of
the masses of $^{64}$Ge\cite{Cla07,Sch07}, $^{65}$As\cite{Tu11} (first mentioned in Wallace and Woosley\cite{Wal84} in connection
with accreting neutron stars),$^{68}$Se\cite{Cla04, Char05}, $^{69}$Br\cite{Rog11}, and $^{72}$Kr\cite{Rod04}; measurements of excited states in $^{66}$Se\cite{Obe11,Ruo13}; and measurements to improve, e.g., the $^{23}$Al($p,\gamma$)$^{24}$Si\cite{Ban12} and $^{30}$S($\alpha,p$)$^{33}$Cl\cite{Dei11} rates have helped to address key uncertainties around waiting points in model calculations of type I XRBs.  As with classical nova explosions, however, detailed results from sensitivity studies coupling a hydrodynamic model with the required network (up to at least $\approx$Te\cite{Sch01,Koi04}) have not been reported, although partial and preliminary results have been presented\cite{AmtPhD,Amt06} (as mentioned above).  Such studies are needed to directly gauge the impact of nuclear physics uncertainties on observable properties of type I X-ray bursts.  As well, additional tests with different hydrodynamic models are needed to clarify possible discrepancies in predictions using different $^{15}$O($\alpha,\gamma$)$^{19}$Ne and $^{18}$Ne($\alpha,p$)$^{21}$Na rates (see above).  Moreover, since ``stellar" (temperature- and density-dependent) weak interaction rates should ideally be used rather than ``laboratory" rates, we encourage the development of improved, consistent treatments for calculating stellar weak interaction rates for all isotopes in a typical XRB network.  (Following a few recent measurements\cite{Baz08,Sto09}, we note that ``laboratory" weak interaction rates are available for most species involved in XRBs; as well, beta-delayed proton decay seems to have little influence on model predictions, at least in the mass range A = $92 - 101$\cite{Lor12}.)  Stellar weak interaction rates have previously been computed for different ranges of nuclei (e.g., A = $1-39$\cite{Oda94}; A = $45-65$\cite{LMP01}; A = $21-60$\cite{FFN}; A = $65-80$\cite{PF03}).  For some species, these rates may differ significantly from laboratory weak rates (see, e.g., Sarriguen\cite{Sar11}) which may dramatically affect predictions of XRB light curves\cite{Woo04}.

\section{Outlook}

When feasible, modelers should work to evolve multidimensional hydrodynamic model calculations of both classical nova explosions and type I X-ray bursts from the accretion stage through the explosion and ejection (for novae) stages.  From the point of view of the associated nucleosynthesis, however, the limited nuclear networks coupled to these computationally-demanding simulations assures the endurance of results from 1-D models for now.  

We have attempted to summarize reactions and nuclear masses that have been identified in previous studies as having uncertainties that significantly affect model predictions of classical novae and type I X-ray bursts.  Nevertheless, new comprehensive sensitivity studies are needed, ideally using 1-D hydrodynamic models coupled to updated networks with current nuclear physics uncertainties. These studies should also focus upon resolving possible discrepancies observed between current XRB models, such as
the impact of uncertainties in the $^{15}$O($\alpha,\gamma$) or $^{18}$Ne($\alpha,p$) rates on predicted light curves or discerning the exact role of the metallicity of the accreted material on observable predictions. Detailed calculations using, e.g., sufficiently large numbers of type I X-ray bursts to ensure convergence of the calculations, would help to shed light on these issues.  In the meantime, experimentalists should build further upon recent accomplishments to fully characterize the rates of reactions such as $^{18}$F($p,\alpha$), $^{23}$Mg($p,\gamma$), $^{25}$Al($p,\gamma$) and $^{30}$P($p,\gamma$) for nova explosions and $^{14}$O($\alpha,p$), $^{15}$O($\alpha,\gamma$), $^{22}$Mg($\alpha,p$), $^{23}$Al($p,\gamma$), $^{30}$S($\alpha,p$), $^{59}$Cu($p,\gamma$) and $^{65}$As($p,\gamma$) for XRBs.  Consistent treatments for calculating stellar weak rates
for all isotopes in a typical XRB network are also needed.

New observatories, such as the recently-launched NuSTAR\cite{NUSTAR} and proposed LOFT\cite{LOFT} missions, hold promise for the identification of absorption features from XRBs. Such results
could provide a needed direct constraint on nucleosynthesis in these environments. As well,
prospects for the ejection of nuclear-processed material by radiation-driven winds from XRBs
still need to be evaluated through detailed models. For novae, UV observations had often been used in the past
to determine abundances of ejected material, and the WSO--UV\cite{WSOUV}
project will help to further advance such studies. High
resolution X-ray spectra of nova ejecta have been obtained by e.g., Chandra\cite{CHANDRA} and XMM-Newton\cite{XMM},
however improvements in expanding atmosphere models are necessary before abundances can be
reliably extracted from these spectra.

% If in two-column mode, this environment will change to single-column format so that long equations can be displayed. 
%\begin{widetext}
%$$\mbox{put long equation here}$$
%\end{widetext}

% Figures should be put into the text as floats. 
% Use the graphics or graphicx packages (distributed with LaTeX2e).
% Put the label that you will use with \ref{} command in the braces of the \label{} command.
%
% \begin{figure}
% \end{figure}

% \begin{table}
% \end{tabular}
% \end{table}

\begin{acknowledgments}
% Put your acknowledgments here.
We thank K. Chipps and M. Bertolli for the kind invitation to submit this article.  This work was supported by the Spanish MICINN under Grants No. AYA2010-15685 and No. EUI2009-04167, by the E. U. FEDER funds, and by the ESF EUROCORES Program EuroGENESIS.
\end{acknowledgments}

%\bibliography{AIPADref4}

%%%%%%%%%%%%%%%19dec2013..............changed

%

\end{document}